\documentclass[10pt]{IEEEtran}
\usepackage[LGR,T1]{fontenc}
\usepackage[latin9]{inputenc}
\usepackage{color}
\usepackage{mathrsfs}
\usepackage{amsmath}
\usepackage{amsthm}
\usepackage{amssymb}
\usepackage{graphicx}
\makeatletter

\providecommand{\tabularnewline}{\\}

\theoremstyle{plain}
\newtheorem{thm}{\protect\theoremname}
\theoremstyle{plain}
\newtheorem{lem}{\protect\lemmaname}
\theoremstyle{remark}
\newtheorem{rem}{\protect\remarkname}
\theoremstyle{plain}
\newtheorem{cor}{\protect\corollaryname}
\theoremstyle{plain}
\newtheorem{prop}{\protect\propositionname}

\usepackage{cite}
  \providecommand{\lemmaname}{Lemma}
  \providecommand{\remarkname}{Remark}
\providecommand{\theoremname}{Theorem}

\allowdisplaybreaks[1]
\providecommand{\corollaryname}{Corollary}
\providecommand{\lemmaname}{Lemma}
\providecommand{\propositionname}{Proposition}
\providecommand{\remarkname}{Remark}
\providecommand{\theoremname}{Theorem}

\makeatother

\begin{document}
\title{On Non-Interactive Simulation of Binary Random Variables}
\author{Lei Yu and Vincent Y. F. Tan, \IEEEmembership{Senior Member,~IEEE}
\thanks{ This work was supported by a Singapore National Research Foundation
(NRF) Fellowship (R-263-000-D02-281) and a Singapore Ministry of Education Tier 2 grant (R-263-000-C83-112). } \thanks{ L. Yu is with the Department of Electrical Engineering and Computer
Sciences, University of California, Berkeley, CA, USA, 94720 (e-mail:
leiyu.scholar@outlook.com). V.~Y.~F.~Tan is with the Department of Electrical and Computer Engineering and the Department of Mathematics, NUS, Singapore 119076 (e-mail: vtan@nus.edu.sg).} \thanks{ Communicated by A. Gohari, Associate Editor at Large. } \thanks{Copyright (c) 2021 IEEE. Personal use of this material is permitted. However, permission to use this material for any other purposes must be obtained from the IEEE by sending a request to pubs-permissions@ieee.org.} }
\maketitle
\begin{abstract}
We leverage proof techniques from discrete Fourier analysis and an
existing result in coding theory to derive new bounds for the problem
of non-interactive simulation of binary random variables. Previous
bounds in the literature were derived by applying data processing
inequalities concerning maximal correlation or hypercontractivity.
We show that our bounds are sharp in some regimes. Indeed, for a specific
instance of the problem parameters, our main result resolves an open
problem posed by E.~Mossel in 2017. As by-products of our analyses,
various new properties of the average distance and distance enumerator
of binary block codes are established. 
\end{abstract}


\section{\label{sec:Introduction}Introduction}

Given a joint distribution $P_{X,Y}$, assume that $(\mathbf{X},\mathbf{Y})\sim P_{\mathbf{X},\mathbf{Y}}:=P_{X,Y}^{n}$
(i.e., $(\mathbf{X},\mathbf{Y})$ are $n$ i.i.d.\ copies of $\left(X,Y\right)\sim P_{X,Y}$)
is a pair of correlated memoryless sources and $(U,V)$ are random
variables such that $U-\mathbf{X}-\mathbf{Y}-V$ forms a Markov chain.\footnote{We say $U,\mathbf{X},\mathbf{Y},V$ forms a Markov chain, denoted
as $U-\mathbf{X}-\mathbf{Y}-V$, if the joint distribution of $U,\mathbf{X},\mathbf{Y},V$
can be written as $P_{U|\mathbf{X}}P_{\mathbf{X},\mathbf{Y}}P_{V|\mathbf{Y}}$.} A natural question arises: What are the possible joint distributions
of $(U,V)$? This problem is termed the \emph{non-interactive simulation
(NIS) problem} \emph{ }\cite{kamath2016non,witsenhausen1975sequences}.
In this paper, we restrict $X,Y,U,V$ to be Boolean random variables
(more precisely, random signs) taking values in $\{-1,1\}$ and $P_{X,Y}$
to be a Boolean symmetric distribution (also known as a doubly symmetric
binary source), i.e., 
\begin{align}
 & \begin{array}{c}
\qquad\qquad\qquad-1\qquad1\\
P_{X,Y}=\begin{array}{c}
-1\\
1
\end{array}\left[\begin{array}{cc}
\frac{1+\rho}{4} & \frac{1-\rho}{4}\vspace{.07in}\\
\frac{1-\rho}{4} & \frac{1+\rho}{4}
\end{array}\right]
\end{array}\label{eq:-34}
\end{align}
with $\rho\in[-1,1]$. Since the joint distribution of the random
variables $U$ and $V$ is entirely determined by the triple of scalars
$\left(\mathbb{P}\left(U=1\right),\mathbb{P}\left(V=1\right),\mathbb{P}\left(U=V=1\right)\right)$,
the closure of the set of the possible joint distributions of $(U,V)$
is determined by the following two quantities: 
\[
\overline{\Theta}\left(a,b\right):=\lim_{n\to\infty}\overline{\Theta}_{n}\left(a,b\right)\quad\textrm{and }\quad\underline{\Theta}\left(a,b\right):=\lim_{n\to\infty}\underline{\Theta}_{n}\left(a,b\right)
\]
where 
\begin{align}
\overline{\Theta}_{n}\left(a,b\right) & :=\max_{\substack{U,V:U-\mathbf{X}-\mathbf{Y}-V\\
\mathbb{P}\left(U=1\right)=a,\\
\mathbb{P}\left(V=1\right)=b
}
}\mathbb{P}\left(U=V=1\right)\label{eq:-7}
\end{align}
and $\underline{\Theta}_{n}\left(a,b\right)$ is defined similarly
but with the maximization replaced by a minimization. The limits here
exist, since $\overline{\Theta}_{n}\left(a,b\right)$ and $\underline{\Theta}_{n}\left(a,b\right)$
are respectively non-increasing and non-decreasing in $n$. Here the
fact that $\overline{\Theta}_{n}\left(a,b\right)$ and $\underline{\Theta}_{n}\left(a,b\right)$
are monotone in $n$ follows from the fact that any pair $(U,V)$
admissible for blocklength $n$ remain admissible for blocklength
$n+1$. Furthermore, by replacing $U$ in \eqref{eq:-7} with $-U$,
it follows that $\underline{\Theta}\left(a,b\right)=b-\overline{\Theta}\left(1-a,b\right)$.
Hence the closure of the set of distributions of $(U,V)$ is determined
only by $\overline{\Theta}\left(a,b\right)$ with $\left(a,b\right)\in\left[0,1\right]^{2}$
(or by $\underline{\Theta}\left(a,b\right)$ and $\overline{\Theta}\left(a,b\right)$
with $\left(a,b\right)\in\left[0,\frac{1}{2}\right]^{2}$). Although
in the definitions of $\underline{\Theta}\left(a,b\right)$ and $\overline{\Theta}\left(a,b\right)$,
$U,V$ are respectively generated by $\mathbf{X},\mathbf{Y}$ through
stochastic maps, in Section \ref{subsec:Non-Interactive-Simulation},
we show that stochastic maps can be replaced by deterministic ones
(i.e., $U=f\left(\mathbf{X}\right)$ and $V=g\left(\mathbf{Y}\right)$),
without affecting the values of $\underline{\Theta}\left(a,b\right)$
and $\overline{\Theta}\left(a,b\right)$.

\subsection{Motivations for the NIS Problem}

The study of the NIS problem is motivated by several applications
in information theory, cryptography, and stochastic control; see~\cite{witsenhausen1975sequences,kamath2016non,yang2007possibility}
for example. The quest to determine $\overline{\Theta}\left(a,b\right)$
or $\underline{\Theta}\left(a,b\right)$ dates back to Witsenhausen's
seminal paper \cite{witsenhausen1975sequences} published in 1975,
in which it was termed the \emph{binary decision problem}. Witsenhausen's
motivation for studying such a problem stemmed from a related information-theoretic
problem---the G\'acs-K\"orner common information problem \cite{gacs1973common}.
In \cite{witsenhausen1975sequences}, Witsenhausen applied the tensorization
property of the maximal correction (introduced by Hirschfeld~\cite{hirschfeld1935connection}
and Gebelein~\cite{gebelein1941statistische}) to provide an upper
bound for $\overline{\Theta}\left(a,b\right)$, which in turn implies
a converse result for the G\'acs-K\"orner common information. In particular,
Witsenhausen addressed the case $a=b=\frac{1}{2}$ by showing that
$\overline{\Theta}\left(\frac{1}{2},\frac{1}{2}\right)$ is attained
by a sequence of dictator functions (i.e., $\mathbf{x}\mapsto x_{i}$
for $1\le i\le n$). However, the determination of $\overline{\Theta}\left(a,b\right)$
(or $\underline{\Theta}\left(a,b\right)$) has been open for all $\left(a,b\right)\in\left(0,1\right)^{2}\backslash\left\{ \left(\frac{1}{2},\frac{1}{2}\right)\right\} $.
Mossel et al. \cite{mossel2006non} and O'Donnell \cite{O'Donnell14analysisof} 
as well as  Kamath and Anantharam \cite{kamath2016non} used the forward
and reverse hypercontractivity to obtain bounds for the NIS problem;
in particular, Kamath and Anantharam's bounds outperform Witsenhausen's
maximal correction bounds. In fact, E.~Mossel posed the determination
of $\overline{\Theta}\left(\frac{1}{4},\frac{1}{4}\right)$ and $\underline{\Theta}\left(\frac{1}{4},\frac{1}{4}\right)$
as open problems in 2017~\cite{Mossel2017}. In this paper, we fully
resolve one of these two open problems, namely, the determination
of $\overline{\Theta}\left(\frac{1}{4},\frac{1}{4}\right)$. We also
provide new bounds for other values of $a$ and $b$. We show that
our bounds, which are derived based on discrete Fourier analysis,
outperform existing ones \cite{witsenhausen1975sequences,kamath2016non}
in some regimes.

\subsection{Connections to Other Problems}

In the literature, the determination of $\overline{\Theta}_{n}\left(a,b\right)$
is also termed the \emph{non-interactive correlation distillation
(NICD) problem}{} \cite{yang2007possibility,mossel2005coin,mossel2006non}.
Deterministic protocols were considered by Mossel, O'Donnell, and
other coauthors~\cite{mossel2005coin,mossel2006non}; while stochastic
protocols were permitted and studied in Yang's work \cite{yang2007possibility}.
(As mentioned previously, these two scenarios are equivalent.) Moreover,
the $k$-terminal version of the NICD problem with $k\ge3$ was also
studied by Mossel and O'Donnell \cite{mossel2005coin}. In this version
of the problem, given noise-corrupted versions $\mathbf{X}^{(i)},1\le i\le k$
of a source $\mathbf{X}$, it is required to maximize the collision
probability $\mathbb{P}\left(f_{1}(\mathbf{X}^{(1)})=\ldots=f_{k}(\mathbf{X}^{(k)})\right)$
over all \emph{balanced} Boolean functions $f_{1},\ldots,f_{k}$.
The authors showed that when the number of parties $k$ is $3$, the
optimal functions are dictator functions \cite{mossel2005coin}. This
extends Witsenhausen's result from the $2$-terminal case to the $3$-terminal
case.

In the study of the NICD problem, it is usually assumed that $a=b$.
It is not difficult to show that if $a=b$ are dyadic rationals, then
there is a deterministic symmetric protocol $\left(f,f\right)$ that
attains $\overline{\Theta}_{n}\left(a,a\right)$; see Lemma \ref{lem:D}
in Section \ref{sec:Basic-Properties-of}. In other words, if $a$
is a dyadic rational, 
\[
\overline{\Theta}_{n}\left(a,a\right)=\max_{\substack{f:\mathbb{P}\left(f(\mathbf{X})=1\right)=a}
}\mathbb{P}\left(f(\mathbf{X})=f(\mathbf{Y})=1\right).
\]
The determination of this quantity is also called \emph{noise stability
problem}, a problem introduced by Benjamini, Kalai, and Schramm \cite{benjamini1999noise};
also see a short survey on \cite[p. 68]{O'Donnell14analysisof}. Hence
the NICD problem (or the maximization part of the NIS problem) can
be seen as a generalization of the noise stability problem to the
case in which $a$ is not necessarily equal to $b$. Furthermore,
the NIS problem is also related to Courtade and Kumar's conjecture
on the most informative Boolean function \cite{courtade2014boolean}
(a weaker version of which was solved by Pichler, Piantanida, and
Matz \cite{pichler2018dictator}) and the problem studied by Li and
M\'edard \cite{li2019boolean} concerning the determination of Boolean
functions that have the maximum $\alpha$-moment. A variant of NIS
problem was also considered by Ordentlich, Polyanskiy and Shayevitz~\cite{ordentlich2020note}.
In \cite{ordentlich2020note}, the authors considered the scenario
in which $a=a_{n}$ and $b=b_{n}$ are allowed to vary with $n$ and
both sequences vanish exponentially as $n\to\infty$. The optimal
exponent of this variant for the case $a_{n}=b_{n}$ is implied by
an enhanced version of the hypercontractivity inequality provided
in \cite{kirshner2019moment}. The optimal exponents for arbitrary
distributions with \emph{finite}{} alphabets and for both symmetric
and asymmetric cases have recently been completely characterized by
Yu, Anantharam, and Chen \cite{yu2021Graphs}.

It can also be verified that the NIS problem is equivalent to a coding-theoretic
problem concerning the {\em maximizing the distance enumerator}
between two codes. Moreover, the maximum distance enumerator problem
is closely related to the {\em minimum average distance} problem
proposed in \cite{ahlswede1977contributions}. To describe this connection,
we recap a few definitions. A subset $C$ of $\{-1,1\}^{n}$ with
size $M$ is called a {\em binary $(n,M)$-code}. The {\em average
distance} of $C$ is defined as the average Hamming distance of every
pair of codewords in $C$. Ahlswede and Katona~\cite{ahlswede1977contributions}
asked a question of determining the minimum of the average distance
of $C$ over all sets $C\subseteq\{-1,1\}^{n}$ of a given cardinality
$M$. Alth\"ofer and Sillke \cite{althofer1992average}, Fu, Xia, together
with other authors \cite{shutao1998average,fu1997expectation,fu1999hamming,fu2001minimum},
as well as Mounits \cite{mounits2008lower}, studied this problem
and proved various bounds on the minimum average distance, which are
sharp in certain regimes with ``large'' size of code (e.g., $M=2^{n-1}$
or $2^{n-2}$). Ahlswede and Alth\"ofer~\cite{ahlswede1994asymptotic}
considered the case in which the size of code increases exponentially
in $n$ (with the exponent strictly sandwiched between $0$ and $1$),
and showed that Hamming spheres are asymptotically optimal solutions
to the Ahlswede-Katona problem as $n\to\infty$. Mounits \cite{mounits2008lower}
studied the case in which the size of the code is ``small'' (in
fact, linear in $n$). In this paper, we consider two variants of
the Ahlswede-Katona's problem. The average distance between two possibly
different codes $A,B\subseteq\{-1,1\}^{n}$ is defined as the average
Hamming distance between every pair of codewords in $A$ and $B$.
The distance enumerator between $A,B\subseteq\{-1,1\}^{n}$ is defined
as the generating function of the distance distribution between $A$
and $B$. For $1\le M,N\le2^{n}$, we are interested in determining
the minima and the maxima of the average distance and the distance
enumerator between $A$ and $B$ over all sets $A,B\subseteq\{-1,1\}^{n}$
of given cardinalities $M,N$ respectively. The average distance problem
considered here extends Ahlswede and Katona's version by allowing
the codewords to come from two different codes. On the other hand,
by a simple approximation argument (cf. Lemma \ref{lem:Approximation}
in Subsection \ref{subsec:Non-Interactive-Simulation}), it can be
seen that the distance enumerator problem is asymptotically equivalent
to the NIS problem as $n\to\infty$. The average distance and the
distance enumerator are closely related, and hence results on the
average distance problem can be applied to the distance enumerator
problem and the NIS problem, and vice verse.

\subsection{Our Main Contributions}

In this paper, we study properties of the average distance and distance
enumerator for the case of two binary codes $A$ and $B$. Then by
combining the use of discrete Fourier analysis techniques and coding-theoretic
results concerning the minimum average distance problem, we derive
bounds on the distance enumerator. By leveraging the equivalence between
the distance enumerator and the NIS problem, we show that bounds on
the former imply some new and nontrivial results on the latter. Moreover,
we show that the bounds are sharp for the special case in which $a=b=\frac{1}{4}$.
The bounds are tighter than the hypercontractivity bound provided
by Kamath and Anantharam \cite{kamath2016non} in some other regimes,
but not in all. For the case in which $a=b=\frac{1}{4}$, our result
answers an open problem posed by E.~Mossel in 2017~\cite{Mossel2017}.

\subsection{Paper Organization}

This paper is organized as follows. In Section \ref{sec:defs}, we
introduce the definitions of coding-theoretic quantities, including
the distance distribution, the average distance, and the distance
enumerator. In this section, we also provide the formulation of the
NIS problem. In Section \ref{sec:Basic-Properties-of}, we study properties
of the average distance and the distance enumerator. In Section \ref{sec:nis},
these properties are applied to derive bounds on the distance enumerator,
or equivalently, bounds for the NIS problem. Finally, in Section \ref{sec:conc},
we provide concluding remarks.

\section{Definitions and Preliminaries}

\label{sec:defs}

\subsection{Distance Distributions}

For two subsets of the Boolean hypercube (termed \emph{codes}) $A,B\subseteq\{-1,1\}^{n}$,
the \emph{distance distribution} between $A$ and $B$ is a probability
mass function $P^{\left(A,B\right)}$ such that for $i\in\{0,1,\ldots,n\}$,
\[
P^{\left(A,B\right)}(i):=\frac{1}{|A||B|}\left|\left\{ \left(\mathbf{x},\mathbf{x}'\right)\in A\times B:d_{\mathrm{H}}\left(\mathbf{x},\mathbf{x}'\right)=i\right\} \right|,
\]
where $d_{\mathrm{H}}\left(\mathbf{x},\mathbf{x}'\right):=\left|\left\{ i:\:x_{i}\neq x'_{i}\right\} \right|$
denotes the Hamming distance between vectors $\mathbf{x},\mathbf{x}'$
(i.e., the number of components of $\mathbf{x}$ and $\mathbf{x}'$
that differ). It is clear that $P^{\left(A,B\right)}(0)=\frac{\left|A\cap B\right|}{|A||B|}$,
$\sum_{i=0}^{n}P^{\left(A,B\right)}(i)=1$, and $P^{\left(A,B\right)}(i)\ge0$
for $i\in\{0,1,\ldots,n\}$.

Define the \emph{$k$-th distance moment} and the \emph{distance enumerator}
between $A,B\subseteq\{-1,1\}^{n}$ respectively as 
\begin{align*}
D_{k}\left(A,B\right) & :=\frac{1}{|A||B|}\sum_{\mathbf{x}\in A}\sum_{\mathbf{x}'\in B}d_{\mathrm{H}}^{k}\left(\mathbf{x},\mathbf{x}'\right)\\
 & =\sum_{i=0}^{n}P^{\left(A,B\right)}(i)\cdot i^{k},
\end{align*}
and 
\begin{align*}
\Gamma_{z}\left(A,B\right) & :=\frac{1}{|A||B|}\sum_{\mathbf{x}\in A}\sum_{\mathbf{x}'\in B}z^{d_{\mathrm{H}}\left(\mathbf{x},\mathbf{x}'\right)}\\
 & =\sum_{i=0}^{n}P^{\left(A,B\right)}(i)\cdot z^{i},\quad z\ge0.
\end{align*}
Clearly, $D_{k}\left(A,B\right)$ and $\Gamma_{z}\left(A,B\right)$
are respectively the $k$-th moment and the generating function of
$P^{\left(A,B\right)}$, and hence they are closely related. For $k=1$,
$D\left(A,B\right):=D_{1}\left(A,B\right)$ corresponds to the {\em
average distance}.

The \emph{dual distance distribution} between codes $A,B$ is defined
by 
\begin{align}
Q^{\left(A,B\right)}(i) & :=\frac{1}{|A||B|}\sum_{\mathbf{u}\in\{0,1\}^{n}:w_{\mathrm{H}}(\mathbf{u})=i}\left(\sum_{\mathbf{x}\in\frac{A+1}{2}}\left(-1\right)^{\left\langle \mathbf{u},\mathbf{x}\right\rangle }\right)\nonumber \\
 & \qquad\times\left(\sum_{\mathbf{x}\in\frac{B+1}{2}}\left(-1\right)^{\left\langle \mathbf{u},\mathbf{x}\right\rangle }\right),\;i=0,1,\ldots,n,\label{eq:-30}
\end{align}
where $w_{\mathrm{H}}(\mathbf{u}):=d_{\mathrm{H}}\left(\mathbf{u},\mathbf{0}\right)$
denotes the Hamming weight (i.e., the number of nonzero components)
of a vector $\mathbf{u}$, and $\left\langle \mathbf{u},\mathbf{x}\right\rangle :=\left(\sum_{i=1}^{n}u_{i}x_{i}\right)\,\mathrm{mod}\,2$
denotes the inner product of vectors $\mathbf{u},\mathbf{x}\in\mathbb{F}_{2}^{n}$.
Clearly, $Q^{\left(A,B\right)}(0)=1$. Note that $Q^{\left(A,B\right)}(i)$
can be negative for some $i$, and $\sum_{i=0}^{n}Q^{\left(A,B\right)}(i)$
can be smaller or larger than $1$. Hence in general, $Q^{\left(A,B\right)}$
is not a probability mass function.

The \emph{dual distance enumerator} between $A,B$ is defined as 
\begin{equation}
\Pi_{z}\left(A,B\right):=\sum_{i=0}^{n}Q^{\left(A,B\right)}(i)\cdot z^{i},\quad z\ge0.\label{eq:-51}
\end{equation}
It is easy to verify that the following MacWilliams--Delsarte identities
hold. 
\begin{align}
\Pi_{z}\left(A,B\right) & =\left(1+z\right)^{n}\Gamma_{\frac{1-z}{1+z}}\left(A,B\right)\label{eq:-22}\\
\Gamma_{z}\left(A,B\right) & =\left(\frac{1+z}{2}\right)^{n}\Pi_{\frac{1-z}{1+z}}\left(A,B\right).\label{eq:-23}
\end{align}
If $A=B$, then $Q^{\left(A,A\right)}(i)\ge0$, and by \eqref{eq:-22},
$\sum_{i=0}^{n}Q^{\left(A,A\right)}(i)=\frac{2^{n}}{|A|}$. Hence
for this case, $\frac{|A|}{2^{n}}Q^{\left(A,A\right)}(\cdot)$ is
a probability mass function. Furthermore, the MacWilliams--Delsarte
identities for the case $A=B$ can be found in \cite{shutao1998average,macwilliams1977theory}.

For brevity, we utilize the following abbreviations: $P^{\left(A\right)}:=P^{\left(A,A\right)}$,
$Q^{\left(A\right)}:=Q^{\left(A,A\right)}$, $D_{k}\left(A\right):=D_{k}\left(A,A\right)$,
$\Gamma_{z}\left(A\right):=\Gamma_{z}\left(A,A\right)$ and $\Pi_{z}\left(A\right):=\Pi_{z}\left(A,A\right)$.

\subsection{Basics of Discrete Fourier Analysis}

We now provide a brief primer on the discrete Fourier analysis needed
in this paper. We also discuss its relationship to the dual distance
distribution. Define $[n]:=\left\{ 1,2,\ldots,n\right\} $. Given
the basis $\chi_{S}(\mathbf{x}):=\prod_{i\in S}x_{i}$ for $S\subseteq[n]$,
for a Boolean function $f:\{-1,1\}^{n}\to\{-1,1\}$, define its {\em
Fourier coefficients} as 
\begin{align}
 & \hat{f}_{S}:=\mathbb{E}_{\mathbf{X}\sim\mathrm{Unif}\left\{ -1,1\right\} ^{n}}[f(\mathbf{X})\chi_{S}(\mathbf{X})],\quad S\subseteq[n].\label{eq:-33}
\end{align}
Then the {\em Fourier expansion} of a Boolean function $f$ (cf.
\cite[Eqn.~(1.6)]{O'Donnell14analysisof}) is 
\begin{align*}
 & f(\mathbf{x})=\sum_{S\subseteq[n]}\hat{f}_{S}\chi_{S}(\mathbf{x}).
\end{align*}
By definition, it is easy to verify that for two codes $A,B\subseteq\{-1,1\}^{n}$
with their $\{-1,1\}$-valued indicator functions denoted as $f=2\cdot\mathbf{1}_{A}-1,g=2\cdot\mathbf{1}_{B}-1$,
the dual distance distribution $Q^{\left(A,B\right)}$ and the Fourier
coefficients $\hat{f}_{S},\hat{g}_{S}$ satisfy the following relationship:\footnote{Note that if $f=\mathbf{1}_{A},g=\mathbf{1}_{B}$ are $\{0,1\}$-valued
indicator functions, then \eqref{eq:-52} reduces to the simpler relation:
\[
Q^{\left(A,B\right)}(k)=\frac{1}{ab}\sum_{S:|S|=k}\hat{f}_{S}\hat{g}_{S},\quad0\le k\le n
\]
} 
\begin{equation}
Q^{\left(A,B\right)}(k)=\begin{cases}
1, & k=0\\
\frac{1}{4ab}\sum_{S:|S|=k}\hat{f}_{S}\hat{g}_{S}, & 1\le k\le n
\end{cases}.\label{eq:-52}
\end{equation} 

\subsection{\label{subsec:Non-Interactive-Simulation}Non-Interactive Simulation }

Let $\left(X,Y\right)$ be a pair of binary random variables on $\{-1,1\}$
with the distribution given in \eqref{eq:-34}, where $\rho\in[-1,1]$
denotes the correlation between $X$ and $Y$. Assume $\left(\mathbf{X},\mathbf{Y}\right)$
are $n$ i.i.d. copies of $\left(X,Y\right)$. Then we focus on the
following NIS problem: Given $a,b\in[0,1]$, what is the possible
range of the collision probability $\mathbb{P}\left(U=V\right)$ for
all Boolean random variables $U,V$ (or equivalently, conditional
Boolean distributions $P_{U|\mathbf{X}},P_{V|\mathbf{Y}}$) such that
$U-\mathbf{X}-\mathbf{Y}-V$ and $\mathbb{P}\left(U=1\right)=a,\mathbb{P}\left(V=1\right)=b$?
Obviously, $\mathbb{P}\left(U=V\right)=1+2\mathbb{P}\left(U=V=1\right)-a-b$.
Hence it suffices to consider the possible range (i.e., the maximum
and minimum) of $\mathbb{P}\left(U=V=1\right)$. Furthermore, the
maximum and minimum $\mathbb{P}\left(U=V=1\right)$ can be approximated
by replacing the conditional Boolean distributions $P_{U|\mathbf{X}},P_{V|\mathbf{Y}}$
with Boolean functions, as shown by the following lemma. For dyadic
rationals $a_{n}:=\frac{\left\lfloor 2^{n}a\right\rfloor }{2^{n}}$
and $b_{n}:=\frac{\left\lfloor 2^{n}b\right\rfloor }{2^{n}}$ with
$\left\lfloor x\right\rfloor $ denoting the maximum integer not larger
than $x$, define 
\begin{equation}
\Delta_{n}^{+}:=\overline{\Theta}_{n}\left(a,b\right)-\max_{\substack{f,g:\mathbb{P}\left(f(\mathbf{X})=1\right)=a_{n},\\
\mathbb{P}\left(g(\mathbf{Y})=1\right)=b_{n}
}
}\mathbb{P}\left(f(\mathbf{X})=g(\mathbf{Y})=1\right),\label{eq:Delta}
\end{equation}
with $\overline{\Theta}_{n}\left(a,b\right)$ defined in \eqref{eq:-7},
and define $\Delta_{n}^{-}$ as the RHS of \eqref{eq:Delta} with
both maximizations (including the one in the definition of $\overline{\Theta}_{n}\left(a,b\right)$)
replaced by minimizations. 
\begin{lem}
\label{lem:Approximation} It holds that 
\begin{equation}
0\leq\Delta_{n}^{+},\Delta_{n}^{-}\leq a-a_{n}+b-b_{n}\leq2^{-\left(n-1\right)}.\label{eq:BoundDelta}
\end{equation}
In particular, if $n,a,b$ are such that $a=\frac{M}{2^{n}}$ and
$b=\frac{N}{2^{n}}$ with some $M,N\in\mathbb{N}$, then $\Delta_{n}^{+}=\Delta_{n}^{-}=0$,
i.e., the maximum or minimum $\mathbb{P}\left(U=V=1\right)$ is attained
by some Boolean functions $f,g:\{-1,1\}^{n}\to\{-1,1\}$ such that
$U=f(\mathbf{X})$ and $V=g(\mathbf{Y})$. 
\end{lem}
\begin{IEEEproof}
First, we claim that for arbitrary $n\in\mathbb{N}$ and $a,b\in[0,1]$,
the maximum or minimum $\mathbb{P}\left(U=V=1\right)$ over all Boolean
random variables $U,V$ such that $U-\mathbf{X}-\mathbf{Y}-V$ and
$\mathbb{P}\left(U=1\right)=a,\mathbb{P}\left(V=1\right)=b$ is attained
by some pair of conditional Boolean distributions $(P_{U|\mathbf{X}},P_{V|\mathbf{Y}})$
with the following properties: There exists at most one $\mathbf{x}_{0}\in\{-1,1\}^{n}$
such that $0<P_{U|\mathbf{X}}(1|\mathbf{x}_{0})<1$ and there exists
at most one $\mathbf{y}_{0}\in\{-1,1\}^{n}$ such that $0<P_{V|\mathbf{Y}}(1|\mathbf{y}_{0})<1$.
This claim follows from the following argument. First, observe that
$\mathbb{P}\left(U=V=1\right)$ can be written as 
\[
\mathbb{P}(U=V=1)=\sum_{\mathbf{x}\in\{-1,1\}^{n}}\lambda_{\mathbf{x}}P_{U|\mathbf{X}}(1|\mathbf{x})
\]
where 
\[
\lambda_{\mathbf{x}}=\sum_{\mathbf{y}}P_{\mathbf{X,Y}}(\mathbf{x},\mathbf{y})P_{V|\mathbf{Y}}(1|\mathbf{y}).
\]
Hence, for fixed $P_{V|\mathbf{Y}}$, the maximization or minimization
of $\mathbb{P}\left(U=V=1\right)$ can be considered as a linear programming
problem with the decision variable being the vector $\left(P_{U|\mathbf{X}}(1|\mathbf{x})\,:\mathbf{x}\in\{-1,1\}^{n}\right)$
whose components satisfy $0\leq P_{U|\mathbf{X}}(1|\mathbf{x})\leq1$
and $\mathbb{P}\left(U=1\right)=\frac{1}{2^{n}}\sum_{\mathbf{x}\in\{-1,1\}^{n}}P_{U|\mathbf{X}}(1|\mathbf{x})=a$.
These constraints correspond to the intersection of the hyperplane
$\big\{\mathbf{t}\in\mathbb{R}^{2^{n}}:\frac{1}{2^{n}}\sum_{i=1}^{2^{n}}t_{i}=a\big\}$
and the hypercube $[0,1]^{2^{n}}$. This intersection is a convex
polygon. Hence, by the fundamental theorem of linear programming~\cite{Bertsimas97},
the maximum and minimum of $\mathbb{P}\left(U=V=1\right)$ are attained
at the corners (i.e., extreme points) of this polygon. Observe that
each corner point of the polygon lies between two adjacent corner
points of the cube. Hence they are precisely the points $P_{U|\mathbf{X}}(1|\cdot)$
such that $0\leq P_{U|\mathbf{X}}(1|\mathbf{x}_{0})\leq1$ for some
$\mathbf{x}_{0}$ and $P_{U|\mathbf{X}}(1|\mathbf{x})=0$ or $1$
for all $\mathbf{x}\neq\mathbf{x}_{0}$. 

Furthermore, for arbitrary $n\in\mathbb{N}$ and $a,b\in[0,1]$, observe
that the contribution of $P_{U|\mathbf{X}}(1|\mathbf{x}_{0})$ to
the probability $\mathbb{P}\left(U=1\right)$ is $P_{U|\mathbf{X}}(1|\mathbf{x}_{0})P_{\mathbf{X}}(\mathbf{x}_{0})=a-a_{n}$.
The contribution of $P_{U|\mathbf{X}}(1|\mathbf{x}_{0})$ to the probability
$\mathbb{P}\left(U=V=1\right)$ is $P_{U|\mathbf{X}}(1|\mathbf{x}_{0})P_{\mathbf{X}}(\mathbf{x}_{0})P_{V|\mathbf{X}}(1|\mathbf{x}_{0})$
which is lower bounded by $0$ and upper bounded by $P_{U|\mathbf{X}}(1|\mathbf{x}_{0})P_{\mathbf{X}}(\mathbf{x}_{0})=a-a_{n}$.
Choose $f(\mathbf{x})=2P_{U|\mathbf{X}}(1|\mathbf{x})-1$ for $\mathbf{x}\neq\mathbf{x}_{0}$
and $f(\mathbf{x}_{0})=-1$ for $\mathbf{x}_{0}$ such that $0<P_{U|X}(1|\mathbf{x}_{0})<1$
provided such an $\mathbf{x}_{0}$ exists; otherwise, choose $f(\mathbf{x})=2P_{U|\mathbf{X}}(1|\mathbf{x})-1$
for all $\mathbf{x}$. Apply the same argument above to $P_{V|\mathbf{Y}}$,
and choose $g$ in a similar way. This concludes the proof of Lemma
\ref{lem:Approximation}. 
\end{IEEEproof}
From this lemma, we have $\lim_{n\to\infty}\Delta_{n}^{+}=\lim_{n\to\infty}\Delta_{n}^{-}=0$.
Hence in this paper, we restrict the maximization (resp. minimization)
of $\mathbb{P}\left(U=V=1\right)$ over Markov chains $U-\mathbf{X}-\mathbf{Y}-V$
to the maximization (resp. minimization) over pairs of Boolean functions
$f,g:\{-1,1\}^{n}\to\{-1,1\}$ such that $U=f(\mathbf{X})$ and $V=g(\mathbf{Y})$.
That is, we consider the following key question: 
\begin{quote}
Given $a=\frac{M}{2^{n}}$ and $b=\frac{N}{2^{n}}$ for some $M,N\in\mathbb{N}$,
what are the maximum and minimum values of the probability $\mathbb{P}\left(f(\mathbf{X})=g(\mathbf{Y})=1\right)$
over all Boolean functions $f,g:\{-1,1\}^{n}\to\{-1,1\}$ such that
$\mathbb{P}\left(f(\mathbf{X})=1\right)=a$ and $\mathbb{P}\left(g(\mathbf{Y})=1\right)=b$? 
\end{quote}
Denote $A:=\left\{ \mathbf{x}:f(\mathbf{x})=1\right\} $ and $B:=\left\{ \mathbf{x}:g(\mathbf{x})=1\right\} $.
Then for $a=\frac{M}{2^{n}}$ and $b=\frac{N}{2^{n}}$ with some $M,N\in\mathbb{N}$,
$\mathbb{P}\left(f(\mathbf{X})=1\right)=a$ and $\mathbb{P}\left(g(\mathbf{Y})=1\right)=b$
imply that $|A|=M$ and $|B|=N$. By \eqref{eq:-22}, we have 
\begin{align}
 & \mathbb{P}\left(f(\mathbf{X})=g(\mathbf{Y})=1\right)\nonumber \\
 & =\sum_{\mathbf{x}\in A}\sum_{\mathbf{x}'\in B}\left(\frac{1-\rho}{4}\right)^{d_{\mathrm{H}}\left(\mathbf{x},\mathbf{x}'\right)}\left(\frac{1+\rho}{4}\right)^{n-d_{\mathrm{H}}\left(\mathbf{x},\mathbf{x}'\right)}\nonumber \\
 & =\left(\frac{1+\rho}{4}\right)^{n}\sum_{\mathbf{x}\in A}\sum_{\mathbf{x}'\in B}\left(\frac{1-\rho}{1+\rho}\right)^{d_{\mathrm{H}}\left(\mathbf{x},\mathbf{x}'\right)}\nonumber \\
 & =ab\left(1+\rho\right)^{n}\Gamma_{\frac{1-\rho}{1+\rho}}\left(A,B\right)\label{eq:-9}\\
 & =ab\Pi_{\rho}\left(A,B\right)\label{eq:-24}
\end{align}
Hence, solving the Boolean function version of the NIS problem is
equivalent to maximizing and minimizing $\Gamma_{\frac{1-\rho}{1+\rho}}\left(A,B\right)$
or $\Pi_{\rho}\left(A,B\right)$ over all subsets $A,B$ such that
$|A|=M$ and $|B|=N$.

\section{\label{sec:Basic-Properties-of}Basic Properties of Distance Moments
and Distance Enumerators}

For a set $A\subseteq\{-1,1\}^{n}$, define its complement and componentwise
complement respectively as $A^{c}:=\{-1,1\}^{n}\backslash A,$ and
$A^{\star}:=\left\{ -\mathbf{x}:\,\mathbf{x}\in A\right\} $, where
$-\mathbf{x}=\left(-x_{1},-x_{2},\ldots,-x_{n}\right)$ is the componentwise
negation of $\mathbf{x}$. Some salient properties of the average
distances and distance enumerators are summarized as follows: 
\begin{lem}
\label{lem:Property}For $A,B\subseteq\{-1,1\}^{n}$, the following
hold. 
\begin{align}
|A|D\left(A,B\right)+|A^{c}|D\left(A^{c},B\right) & =n2^{n-1};\label{eq:-11}\\
D_{k}\left(A^{\star},B\right)=\sum_{i=0}^{k}\binom{k}{i} & n^{k-i}\left(-1\right)^{i}D_{i}\left(A,B\right);\label{eq:-10}\\
|A|\Gamma_{z}\left(A,B\right)+|A^{c}|\Gamma_{z}\left(A^{c},B\right) & =\left(1+z\right)^{n};\label{eq:-12}\\
\Gamma_{z}\left(A^{\star},B\right) & =z^{n}\Gamma_{\frac{1}{z}}\left(A,B\right).\label{eq:-13}
\end{align}
\end{lem}
Lemma \ref{lem:Property} is straightforward, hence the proof is omitted. 
\begin{rem}
For $k=1$, \eqref{eq:-10} implies that 
\begin{equation}
D\left(A,B\right)+D\left(A^{\star},B\right)=n.\label{eq:-15}
\end{equation}
\end{rem}
\begin{lem}
\label{lem:D} For $A,B\subseteq\{-1,1\}^{n}$, the following hold.
\begin{align}
\left|\frac{n}{2}-D\left(A,B\right)\right| & \leq\sqrt{\left(\frac{n}{2}-D\left(A\right)\right)\left(\frac{n}{2}-D\left(B\right)\right)}\nonumber \\
 & \le\frac{n}{2}-\frac{1}{2}\left(D\left(A\right)+D\left(B\right)\right).\label{eq:-14}
\end{align}
For $0\le z\le1$, 
\begin{equation}
\Gamma_{z}\left(A,B\right)\leq\sqrt{\Gamma_{z}\left(A\right)\Gamma_{z}\left(B\right)}\leq\frac{1}{2}\left(\Gamma_{z}\left(A\right)+\Gamma_{z}\left(B\right)\right);\label{eq:-4}
\end{equation}
and for $z\ge1$, 
\begin{align}
\Gamma_{z}\left(A,B\right) & \leq\sqrt{\Gamma_{z}\left(A^{\star},A\right)\Gamma_{z}\left(B^{\star},B\right)}\nonumber \\
 & \leq\frac{1}{2}\left(\Gamma_{z}\left(A^{\star},A\right)+\Gamma_{z}\left(B^{\star},B\right)\right).\label{eq:-5}
\end{align}
\end{lem}
\begin{IEEEproof}
Denote $f:=2\cdot\mathbf{1}_{A}-1$ and $g:=2\cdot\mathbf{1}_{B}-1$.
Both $\frac{n}{2}-D\left(A,B\right)$ and $\Gamma_{z}\left(A,B\right)$
can be written as bilinear functions of $\hat{f}_{S}$ and $\hat{g}_{S}$.
Then applying the Cauchy--Schwarz inequality, the claims in the lemma
follow. In the following, we provide the detailed proof.

Define 
\begin{align}
 & a:=|A|/2^{n}=\mathbb{P}\left(f(\mathbf{X})=1\right)=\frac{1+\hat{f}_{\emptyset}}{2},\nonumber \\
 & b:=|B|/2^{n}=\mathbb{P}\left(g(\mathbf{X})=1\right)=\frac{1+\hat{g}_{\emptyset}}{2},\nonumber \\
 & \theta_{\rho}:=\frac{1}{4}\sum_{S\subseteq[n]:|S|\ge1}\hat{f}_{S}\hat{g}_{S}\rho^{|S|}.\label{eq:-39}
\end{align}
In analogy to \cite[Plancherel's Theorem and Proposition 1.9]{O'Donnell14analysisof},
the inner product between $f(\mathbf{X})$ and $g(\mathbf{Y})$ satisfies
\begin{equation}
\mathbb{E}\left[f(\mathbf{X})g(\mathbf{Y})\right]=\hat{f}_{\emptyset}\hat{g}_{\emptyset}+4\theta_{\rho}=2\mathbb{P}\left(f(\mathbf{X})=g(\mathbf{Y})\right)-1.\label{eq:-8}
\end{equation}
Defining $\overline{t}:=1-t$ for $t\in[0,1]$ and following by \eqref{eq:-8}
and the fact that 
\[
\mathbb{P}\left(f(\mathbf{X})=g(\mathbf{Y})\right)=1+2\mathbb{P}\left(f(\mathbf{X})=g(\mathbf{Y})=1\right)-a-b,
\]
we can write 
\begin{align}
\mathbb{P}\left(f(\mathbf{X})=g(\mathbf{Y})=1\right) & =ab+\theta_{\rho},\label{eq:-26}
\end{align}
which further implies 
\begin{align}
\mathbb{P}\left(f(\mathbf{X})=-g(\mathbf{Y})=1\right) & =a\overline{b}-\theta_{\rho}\\
\mathbb{P}\left(-f(\mathbf{X})=g(\mathbf{Y})=1\right) & =\overline{a}b-\theta_{\rho}\\
\mathbb{P}\left(f(\mathbf{X})=g(\mathbf{Y})=-1\right) & =\overline{a}\overline{b}+\theta_{\rho}.\label{eq:-31}
\end{align}

Combining \eqref{eq:-26}, \eqref{eq:-24}, and \eqref{eq:-52} yields
the following identity: 
\begin{align}
ab\left(1+\rho\right)^{n}\Gamma_{\frac{1-\rho}{1+\rho}}\left(A,B\right) & =ab\Pi_{\rho}\left(A,B\right)\nonumber \\
 & =ab+\frac{1}{4}\sum_{S:|S|\ge1}\hat{f}_{S}\hat{g}_{S}\rho^{|S|}.\label{eq:-42}
\end{align}
By the Cauchy--Schwarz inequality, we have for any $0\le k\le n$,
\begin{equation}
\left|\sum_{S:|S|=k}\hat{f}_{S}\hat{g}_{S}\right|\leq\sqrt{\left(\sum_{S:|S|=k}\hat{f}_{S}^{2}\right)\left(\sum_{S:|S|=k}\hat{g}_{S}^{2}\right)}.\label{eq:-41}
\end{equation}

Now consider the case of $k=1$. For any $i\in[n]$,

\begin{equation}
\hat{f}_{\left\{ i\right\} }=\frac{1}{2^{n}}\left(\sum_{\mathbf{x}\in A}x_{i}-\sum_{\mathbf{x}\in A^{c}}x_{i}\right)=\frac{2}{2^{n}}\sum_{\mathbf{x}\in A}x_{i},\label{eq:-28}
\end{equation}
since $\sum_{\mathbf{x}\in A}x_{i}+\sum_{\mathbf{x}\in A^{c}}x_{i}=0$.
Similarly, for any $i\in[n]$, 
\begin{equation}
\hat{g}_{\left\{ i\right\} }=\frac{2}{2^{n}}\sum_{\mathbf{x}\in B}x_{i}.\label{eq:-29}
\end{equation}
Therefore, 
\begin{align}
c:=\sum_{S:|S|=1}\hat{f}_{S}\hat{g}_{S} & =\sum_{i=1}^{n}\frac{1}{4^{n-1}}\sum_{\mathbf{x}\in A}x_{i}\sum_{\mathbf{x}'\in B}x_{i}'\label{eq:-32}\\
 & =\frac{1}{4^{n-1}}\sum_{\mathbf{x}\in A}\sum_{\mathbf{x}'\in B}\left(n-2d_{\mathrm{H}}\left(\mathbf{x},\mathbf{x}'\right)\right)\nonumber \\
 & =4ab\left(n-2D\left(A,B\right)\right),\label{eq:-27}
\end{align}
where $a=\frac{|A|}{2^{n}}$ and $b=\frac{|B|}{2^{n}}$. Combining
\eqref{eq:-41} for $k=1$ with \eqref{eq:-27} (and its symmetric
versions with $B$ replaced by $A$ or $A$ replaced by $B$) yields
\begin{align*}
\left|\frac{n}{2}-D\left(A,B\right)\right| & \leq\sqrt{\left(\frac{n}{2}-D\left(A\right)\right)\left(\frac{n}{2}-D\left(B\right)\right)}\\
 & \le\frac{n}{2}-\frac{1}{2}\left(D\left(A\right)+D\left(B\right)\right).
\end{align*}

Now we prove inequality \eqref{eq:-4}. By the Cauchy--Schwarz inequality,
we obtain that for $\rho\in[0,1]$, 
\begin{align}
 & ab+\frac{1}{4}\sum_{S:|S|\ge1}\hat{f}_{S}\hat{g}_{S}\rho^{|S|}\nonumber \\
 & \leq\sqrt{\left(a^{2}+\frac{1}{4}\sum_{S:|S|\ge1}\hat{f}_{S}^{2}\rho^{|S|}\right)\left(b^{2}+\frac{1}{4}\sum_{S:|S|\ge1}\hat{g}_{S}^{2}\rho^{|S|}\right)}.\label{eq:-43}
\end{align}
Combining \eqref{eq:-42} with \eqref{eq:-43} yields 
\begin{align*}
\Gamma_{\frac{1-\rho}{1+\rho}}\left(A,B\right) & \leq\sqrt{\Gamma_{\frac{1-\rho}{1+\rho}}\left(A\right)\Gamma_{\frac{1-\rho}{1+\rho}}\left(B\right)}\\
 & \leq\frac{1}{2}\left(\Gamma_{\frac{1-\rho}{1+\rho}}\left(A\right)+\Gamma_{\frac{1-\rho}{1+\rho}}\left(B\right)\right).
\end{align*}
Setting $z:=\frac{1-\rho}{1+\rho}\in[0,1]$, we obtain that for $0\le z\le1$,
\begin{align*}
\Gamma_{z}\left(A,B\right) & \leq\sqrt{\Gamma_{z}\left(A\right)\Gamma_{z}\left(B\right)}\leq\frac{1}{2}\left(\Gamma_{z}\left(A\right)+\Gamma_{z}\left(B\right)\right).
\end{align*}

Inequality \eqref{eq:-5} follows by combining \eqref{eq:-4} with
\eqref{eq:-13}. 
\end{IEEEproof}
In fact, in the proof of Lemma \ref{lem:D}, we prove the following
inequality: The dual distance distribution between any codes $A$
and $B$ satisfies 
\begin{equation}
\left|Q^{\left(A,B\right)}(k)\right|\leq\sqrt{Q^{\left(A\right)}(k)Q^{\left(B\right)}(k)},\quad0\le k\le n.\label{eq:-44}
\end{equation}
Inequality \eqref{eq:-14} corresponds to inequality \eqref{eq:-44}
with $k=1$. 
\begin{lem}
\label{lem:max-min-D1} For $1\le M\le2^{n}$, 
\begin{align}
\min_{A,B:|A|=|B|=M}D\left(A,B\right) & =\min_{A:|A|=M}D\left(A\right),\label{eq:-16}\\
\max_{A,B:|A|=|B|=M}D\left(A,B\right) & =n-\min_{A:|A|=M}D\left(A\right).\label{eq:-17}
\end{align}
For $0\le z\le1$, 
\begin{equation}
\max_{A,B:|A|=|B|=M}\Gamma_{z}\left(A,B\right)=\max_{A:|A|=M}\Gamma_{z}\left(A\right);\label{eq:-18}
\end{equation}
and for $z\ge1$, 
\begin{equation}
\max_{A,B:|A|=|B|=M}\Gamma_{z}\left(A,B\right)=\max_{A:|A|=M}\Gamma_{z}\left(A^{\star},A\right).\label{eq:-19}
\end{equation}
\end{lem}
%\begin{rem}
%Equation \eqref{eq:-17} implies that for   $1\le M\le2^{n}$, 
%\begin{align}
%\max_{A:|A|=M}D\left(A,A\right) +\min_{A:|A|=M}D\left(A,A\right) & \le n.\label{eq:-17}
%\end{align}
%\end{rem}

\begin{IEEEproof}
By \eqref{eq:-14}, 
\begin{align}
D\left(A,B\right) & \geq\frac{1}{2}\left(D\left(A\right)+D\left(B\right)\right).\label{eq:-14-2}
\end{align}
Minimizing both sides over $A,B$ such that $|A|=|B|=M$, we obtain
\[
\min_{A,B:|A|=|B|=M}D\left(A,B\right)\geq\min_{A:|A|=M}D\left(A\right).
\]
On the other hand, obviously, 
\[
\min_{A,B:|A|=|B|=M}D\left(A,B\right)\leq\min_{A:|A|=M}D\left(A\right).
\]
Hence, \eqref{eq:-16} follows.

Equation \eqref{eq:-17} follow since by \eqref{eq:-15} and \eqref{eq:-16},
\begin{align*}
\max_{A,B:|A|=|B|=M}D\left(A,B\right) & =n-\min_{A,B:|A|=|B|=M}D\left(A^{\star},B\right)\\
 & =n-\min_{A:|A|=M}D\left(A\right),
\end{align*}
Equations \eqref{eq:-18} and \eqref{eq:-19} respectively follow
from \eqref{eq:-4} and \eqref{eq:-5}. 
\end{IEEEproof}
Fu, Wei, and Yeung \cite[Thm. 4]{fu2001minimum} considered the average
distance between codewords in the same set (i.e., restricting $A$
and $B$ to be identical), showed that for $a:=\frac{M}{2^{n}}\le\frac{1}{2}$,
\begin{equation}
\min_{A:|A|=M}D\left(A\right)\geq\frac{n}{2}-\frac{1}{4a},\label{eq:-6}
\end{equation}
by using a linear programming approach. Here equality in \eqref{eq:-6}
holds for $M=2^{n-1}$ or $2^{n-2}$ by setting $A$ to be a subcube
(e.g., $A=\{1\}\times\left\{ -1,1\right\} ^{n-1}$ for $M=2^{n-1}$
and $A=\{1\}^{2}\times\left\{ -1,1\right\} ^{n-2}$ for $M=2^{n-2}$).
Combining this result with Lemmas \ref{lem:Property} and \ref{lem:D}
yields the following bounds on the average distance between two possibly
non-identical sets. 
\begin{lem}
\label{lem:min-max-D2} For $1\le M,N\le2^{n}$, we have 
\begin{align}
\min_{A,B:|A|=M,|B|=N}D\left(A,B\right) & \geq\frac{n}{2}-\frac{\sqrt{\left(a\wedge\overline{a}\right)\left(b\wedge\overline{b}\right)}}{4ab},\label{eq:-20}\\
\max_{A,B:|A|=M,|B|=N}D\left(A,B\right) & \leq\frac{n}{2}+\frac{\sqrt{\left(a\wedge\overline{a}\right)\left(b\wedge\overline{b}\right)}}{4ab},\label{eq:-21}
\end{align}
where $a:=\frac{M}{2^{n}}$, $b:=\frac{N}{2^{n}}$, and $x\wedge y:=\min\left\{ x,y\right\} $.
Equalities in \eqref{eq:-20} and \eqref{eq:-21} hold for $M\in\left\{ 2^{n-1},2^{n-2},3\cdot2^{n-2}\right\} $
and $N=M$ or $2^{n}-M$, by setting $A$ to a subcube, and $B=A$
or $\left(A^{c}\right)^{\star}$ for \eqref{eq:-20}; $B=A$ or $A^{c}$
for \eqref{eq:-21}.  
\end{lem}
\begin{IEEEproof}
By \eqref{eq:-15}, Inequalities \eqref{eq:-20} and \eqref{eq:-21}
are in fact equivalent. Hence, it suffices to prove \eqref{eq:-20}.

Inequality \eqref{eq:-20} for the case $a,b\le\frac{1}{2}$ follows
since by \eqref{eq:-14} and \eqref{eq:-6}, 
\[
\left|\frac{n}{2}-D\left(A,B\right)\right|\leq\sqrt{\left(\frac{n}{2}-D\left(A\right)\right)\left(\frac{n}{2}-D\left(B\right)\right)}\leq\frac{1}{4\sqrt{ab}}.
\]

We now consider the case $b\leq\frac{1}{2}<a$. By \eqref{eq:-11}
and \eqref{eq:-15}, 
\begin{align*}
 & |A|D\left(A,B\right)+|A^{c}|\left(n-D\left(\left(A^{c}\right)^{\star},B\right)\right)\\
 & \qquad=|A|D\left(A,B\right)+|A^{c}|D\left(A^{c},B\right)=n2^{n-1},
\end{align*}
which, combined with \eqref{eq:-20} for the case $a,b\le\frac{1}{2}$,
implies the case $b\leq\frac{1}{2}<a$. By symmetry, the case $b\leq\frac{1}{2}<a$
also holds.

We next consider the case $a,b>\frac{1}{2}$. By \eqref{eq:-11} we
have 
\begin{align*}
|A|D\left(A,B\right)+|A^{c}|D\left(A^{c},B\right) & =n2^{n-1}\\
|B|D\left(B,A^{c}\right)+|B^{c}|D\left(B^{c},A^{c}\right) & =n2^{n-1}.
\end{align*}
Since $D\left(B,A^{c}\right)=D\left(A^{c},B\right)$, we obtain 
\begin{equation}
|A||B|D\left(A,B\right)-|A^{c}||B^{c}|D\left(A^{c},B^{c}\right)=\left(|B|-|A^{c}|\right)n2^{n-1}.\label{eq:-11-1}
\end{equation}
Combining \eqref{eq:-11-1} with inequality \eqref{eq:-20} for the
case $a,b\le\frac{1}{2}$ yields the case $a,b>\frac{1}{2}$. 
\end{IEEEproof}

\section{Non-Interactive Simulation}

\label{sec:nis}

\subsection{Non-Interactive Simulation}

In this section, we derive bounds on $\Gamma_{z}\left(A,B\right)$
and $\Pi_{z}\left(A,B\right)$ for $z\ge0$ and for $A,B$ such that
$|A|=M$ and $|B|=N$. By \eqref{eq:-24}, this is equivalent to bounding
$\mathbb{P}\left(f(\mathbf{X})=g(\mathbf{Y})=1\right)$ for Boolean
functions $f,g$ such that $\mathbb{P}\left(f(\mathbf{X})=1\right)=a$
and $\mathbb{P}\left(g(\mathbf{Y})=1\right)=b$ where $a=\frac{M}{2^{n}}$
and $b=\frac{N}{2^{n}}$. Without loss of generality, we may assume
$0\le a\le b\le\frac{1}{2}$ (or $0\le M\le N\le2^{n-1}$) and $\rho\in[0,1]$.
This is because that if $\rho<0$, then we can then replace $-X$
by $X$; if $a>\frac{1}{2}$ or $b>\frac{1}{2}$, replace $-f$ by
$f$ or $-g$ by $g$. For Boolean functions $f,g$ such that $\mathbb{P}\left(f(\mathbf{X})=1\right)=a$
and $\mathbb{P}\left(g(\mathbf{Y})=1\right)=b$, define 
\[
q:=\mathbb{P}\left(f(\mathbf{X})=g(\mathbf{Y})=1\right).
\]
Then $q$ can be bounded as follows. 
\begin{thm}[Bounds on $q$]
\label{thm:DistanceEnumerator} 
\[
\max\left\{ \Upsilon_{1}^{\mathtt{LB}},\Upsilon_{2}^{\mathtt{LB}}\right\} \leq q\leq\min\left\{ \Upsilon_{1}^{\mathtt{UB}},\Upsilon_{2}^{\mathtt{UB}}\right\} ,
\]
where 
\begin{align*}
\Upsilon_{1}^{\mathtt{LB}} & :=\max\left\{ 0,ab-\frac{\sqrt{ab}}{2}\rho-\frac{ab+\sqrt{a\overline{a}b\overline{b}}}{2}\rho^{2}\right\} \\
\Upsilon_{2}^{\mathtt{LB}} & :=\max\left\{ 0,ab-\frac{\sqrt{ab}}{2}\rho-\frac{a+b-2ab-\sqrt{ab}}{2}\rho^{2}\right\} \\
\Upsilon_{1}^{\mathtt{UB}} & :=\min\left\{ a,ab+\frac{\sqrt{ab}}{2}\rho+\frac{a\overline{b}+\sqrt{a\overline{a}b\overline{b}}-\sqrt{ab}}{2}\rho^{2}\right\} \\
\Upsilon_{2}^{\mathtt{UB}} & :=\sqrt{\theta^{+}(a)\theta^{+}(b)}
\end{align*}
with 
\begin{equation}
\theta^{+}(t):=t^{2}+\frac{t}{2}\rho+\left(\frac{t}{2}-t^{2}\right)\rho^{2}.\label{eqn:theta_plus}
\end{equation}
\end{thm}
\begin{rem}
For $a=b$, $\Upsilon_{1}^{\mathtt{LB}}\le\Upsilon_{2}^{\mathtt{LB}}$,
and for fixed $b$ and sufficiently small $a$, $\Upsilon_{1}^{\mathtt{LB}}\ge\Upsilon_{2}^{\mathtt{LB}}$.
Hence $\Upsilon_{1}^{\mathtt{LB}}\le\Upsilon_{2}^{\mathtt{LB}}$ or
$\Upsilon_{1}^{\mathtt{LB}}\ge\Upsilon_{2}^{\mathtt{LB}}$ does not
always hold. Similarly, $\Upsilon_{1}^{\mathtt{UB}}\le\Upsilon_{2}^{\mathtt{UB}}$
or $\Upsilon_{1}^{\mathtt{UB}}\ge\Upsilon_{2}^{\mathtt{UB}}$ also
does not always hold. 
\end{rem}
\begin{IEEEproof}
The main idea in the proof is as follows. Recall from~\eqref{eq:-26}
that 
\begin{align}
q & =ab+\theta_{\rho}\quad\textrm{ with }\quad\theta_{\rho}=\frac{1}{4}\sum_{k=1}^{n}\left(\sum_{S:|S|=k}\hat{f}_{S}\hat{g}_{S}\right)\rho^{k},\label{eq:-26-1}
\end{align}
which implies that bounding $q$ is equivalent to bounding $\theta_{\rho}$.
By Parseval's theorem and the Cauchy--Schwarz inequality, the sum
of absolute values of coefficients, $\sum_{k=1}^{n}\big|\sum_{S:|S|=k}\hat{f}_{S}\hat{g}_{S}\big|$,
is bounded. On the other hand, recall that (see \eqref{eq:-27}) for
$k=1$, $\sum_{S:|S|=1}\hat{f}_{S}\hat{g}_{S}=4ab\left(n-2D\left(A,B\right)\right)$.
Hence by Lemma \ref{lem:min-max-D2}, we can bound the term $\sum_{S:|S|=1}\hat{f}_{S}\hat{g}_{S}$,
which in turns implies bounds for $q$. Furthermore, in order to further
improve bounds for $q$ in certain cases, we incorporate the ``partition''
technique from Pichler, Piantanida, and Matz \cite{pichler2018dictator}
in our proof. %We next provide
%the detailed proof.

We now provide the details of the proof. By the nonnegativity of the
probabilities given in \eqref{eq:-26}--\eqref{eq:-31}, 
\begin{equation}
-ab\le\theta_{\rho}\le a\overline{b}.\label{eq:}
\end{equation}
As in the proof of Lemma \ref{lem:D}, we write $A:=\left\{ \mathbf{x}:f(\mathbf{x})=1\right\} $
and $B:=\left\{ \mathbf{x}:g(\mathbf{x})=1\right\} $. Then by Lemma
\ref{lem:min-max-D2}, we obtain that 
\begin{align}
\left|\frac{n}{2}-D\left(A,B\right)\right| & \leq\frac{1}{4\sqrt{ab}},\label{eq:-14-1}
\end{align}
since $a,b\le\frac{1}{2}$ as assumed. Hence $c$, as defined in \eqref{eq:-32},
satisfies 
\begin{equation}
|c|\leq2\sqrt{ab}.\label{eq:-38}
\end{equation}

As in the proof of Lemma \ref{lem:D}, we denote the Fourier coefficients
of $f$ and $g$ as $\hat{f}_{S}$ and $\hat{g}_{S}$, respectively;
see~\eqref{eq:-33}. To bound $\theta_{\rho}$, we partition the
set $\{S\subseteq[n]:|S|\ge2\}$ into $\mathcal{P}:=\{S\subseteq[n]:|S|\ge2,\hat{f}_{S}\hat{g}_{S}\ge0\}$
and $\mathcal{N}:=\{S\subseteq[n]:|S|\ge2,\hat{f}_{S}\hat{g}_{S}<0\}$.
This idea for doing this partitioning comes from~\cite{pichler2018dictator}.
We define 
\[
\tau^{+}:=\frac{1}{4}\sum_{S\in\mathcal{P}}\hat{f}_{S}\hat{g}_{S},\quad\mbox{and}\quad\tau^{-}:=\frac{1}{4}\sum_{S\in\mathcal{N}}\hat{f}_{S}\hat{g}_{S}.
\]
Then for $\rho=1$, $\theta_{1}$ can be written as 
\begin{equation}
\theta_{1}=\frac{1}{4}c+\tau^{+}+\tau^{-},\label{eq:-61}
\end{equation}
and by definition, 
\begin{equation}
\frac{1}{4}c\rho+\rho^{2}\tau^{-}\le\theta_{\rho}\le\frac{1}{4}c\rho+\rho^{2}\tau^{+}.\label{eq:-62}
\end{equation}
Hence to bound $\theta_{\rho}$, it suffices to upper bound $\tau^{+}$
and lower bound $\tau^{-}$, which will be done by respectively bounding
$\tau^{+}-\tau^{-}$ and $\tau^{+}+\tau^{-}$. We next do this.

We first apply the Cauchy--Schwarz inequality to bound $\tau^{+}-\tau^{-}$.
\begin{align}
\frac{1}{4}c+\tau^{+}-\tau^{-} & \le\frac{1}{4}\sum_{S:|S|\ge1}|\hat{f}_{S}|\left|\hat{g}_{S}\right|\nonumber \\
 & \le\frac{1}{4}\sqrt{\left(\sum_{S:|S|\ge1}\hat{f}_{S}^{2}\right)\left(\sum_{S:|S|\ge1}\hat{g}_{S}^{2}\right)}\nonumber \\
 & =\frac{1}{4}\sqrt{\left(1-\hat{f}_{\emptyset}^{2}\right)\left(1-\hat{g}_{\emptyset}^{2}\right)}\label{eq:-40}\\
 & =\sqrt{a\overline{a}b\overline{b}},\nonumber 
\end{align}
where \eqref{eq:-40} follows from Parseval's theorem \cite{O'Donnell14analysisof}.
Hence 
\begin{equation}
\tau^{+}-\tau^{-}\le\sqrt{a\overline{a}b\overline{b}}-\frac{1}{4}c.\label{eq:-1}
\end{equation}
We next bound $\tau^{+}+\tau^{-}$. Since $-ab\le\theta_{1}\le a\overline{b}$
(see \eqref{eq:}), by substituting \eqref{eq:-61} in to these inequalities,
we have 
\begin{equation}
-ab-\frac{1}{4}c\le\tau^{+}+\tau^{-}\le a\overline{b}-\frac{1}{4}c.\label{eq:-2}
\end{equation}
Combining \eqref{eq:-1} and \eqref{eq:-2}, we can bound $\tau^{+}$
and $\tau^{-}$ as follows: 
\begin{align}
\tau^{+} & \leq\frac{a\overline{b}+\sqrt{a\overline{a}b\overline{b}}}{2}-\frac{1}{4}c,\qquad\tau^{-}\geq-\frac{ab+\sqrt{a\overline{a}b\overline{b}}}{2}.\label{eq:-63}
\end{align}

Combining the bounds in \eqref{eq:-63} with \eqref{eq:-62} yields
that 
\begin{equation}
\theta_{\rho}\in[\widehat{\theta}_{\rho}^{-},\widehat{\theta}_{\rho}^{+}]\subseteq[\theta_{\rho}^{-},\theta_{\rho}^{+}],\label{eq:-35}
\end{equation}
where, from \eqref{eq:-38}, 
\begin{align*}
\widehat{\theta}_{\rho}^{-} & :=\max\left\{ -ab,\frac{1}{4}c\rho-\frac{ab+\sqrt{a\overline{a}b\overline{b}}}{2}\rho^{2}\right\} \\
 & \geq\max\left\{ -ab,-\frac{\sqrt{ab}}{2}\rho-\frac{ab+\sqrt{a\overline{a}b\overline{b}}}{2}\rho^{2}\right\} =:\theta_{\rho}^{-},\\
\widehat{\theta}_{\rho}^{+} & :=\min\left\{ a\overline{b},\frac{1}{4}c\rho+\left(\frac{a\overline{b}+\sqrt{a\overline{a}b\overline{b}}}{2}-\frac{1}{4}c\right)\rho^{2}\right\} \\
 & \leq\min\left\{ a\overline{b},\frac{\sqrt{ab}}{2}\rho+\frac{a\overline{b}+\sqrt{a\overline{a}b\overline{b}}-\sqrt{ab}}{2}\rho^{2}\right\} =:\theta_{\rho}^{+}.
\end{align*}
Combining \eqref{eq:-35} with \eqref{eq:-26-1} gives us the lower
bound $\Upsilon_{1}^{\mathtt{LB}}$ and upper bound $\Upsilon_{1}^{\mathtt{UB}}$.

We next prove that $q$ is lower bounded by $\Upsilon_{2}^{\mathtt{LB}}$.
Starting from the definition of $\theta_{\rho}$ in \eqref{eq:-39},
we obtain that 
\begin{align}
\theta_{\rho} & \geq\frac{1}{4}c\rho-\frac{1}{4}\sum_{i=2}^{n}\left|\sum_{S:|S|=i}\hat{f}_{S}\hat{g}_{S}\right||\rho|^{i}\nonumber \\
 & \geq-\frac{1}{4}|c|\rho-\frac{1}{4}\left(\sum_{i=2}^{n}\left|\sum_{S:|S|=i}\hat{f}_{S}\hat{g}_{S}\right|\right)\rho^{2}.\label{eq:-3}
\end{align}
We bound the term $\sum_{i=2}^{n}\left|\sum_{S:|S|=i}\hat{f}_{S}\hat{g}_{S}\right|$
by observing that 
\begin{align*}
 & \left(2a-1\right)\left(2b-1\right)+|c|+\sum_{i=2}^{n}\left|\sum_{S:|S|=i}\hat{f}_{S}\hat{g}_{S}\right|\\
 & =\sum_{i=0}^{n}\left|\sum_{S:|S|=i}\hat{f}_{S}\hat{g}_{S}\right|\\
 & \leq\sum_{S\subseteq[n]}|\hat{f}_{S}|\left|\hat{g}_{S}\right|\\
 & \leq\sqrt{\left(\sum_{S\subseteq[n]}\hat{f}_{S}^{2}\right)\left(\sum_{S\subseteq[n]}\hat{g}_{S}^{2}\right)}=1.
\end{align*}
Hence 
\begin{align}
\sum_{i=2}^{n}\left|\sum_{S:|S|=i}\hat{f}_{S}\hat{g}_{S}\right| & \leq1-\left(2a-1\right)\left(2b-1\right)-|c|\nonumber \\
 & =2a+2b-4ab-|c|.\label{eq:-36}
\end{align}
Substituting \eqref{eq:-36} into \eqref{eq:-3}, we obtain 
\begin{align}
\theta_{\rho} & \geq-\frac{1}{4}|c|\rho-\frac{1}{4}\left(2a+2b-4ab-|c|\right)\rho^{2}\nonumber \\
 & \geq-\frac{\sqrt{ab}}{2}\rho-\frac{a+b-2ab-\sqrt{ab}}{2}\rho^{2},\label{eq:-37}
\end{align}
where \eqref{eq:-37} follows since $|c|\leq2\sqrt{ab}$ (see \eqref{eq:-38}).
On the other hand, $\theta_{\rho}\ge-ab$. Hence, 
\[
\theta_{\rho}\geq\max\left\{ -ab,-\frac{\sqrt{ab}}{2}\rho-\frac{a+b-2ab-\sqrt{ab}}{2}\rho^{2}\right\} .
\]
Substituting it into \eqref{eq:-26-1}, we obtain the lower bound
$\Upsilon_{2}^{\mathtt{LB}}$.

We finally prove the upper bound $\Upsilon_{2}^{\mathtt{UB}}$. Denote
\begin{align*}
q_{f} & :=\mathbb{P}\left(f(\mathbf{X})=f(\mathbf{Y})=1\right),\\
q_{g} & :=\mathbb{P}\left(g(\mathbf{X})=g(\mathbf{Y})=1\right).
\end{align*}
From \eqref{eq:-9}, we express $q,q_{f},q_{g}$ in terms of the distance
distributions as follows: 
\begin{align*}
q & =ab\left(1+\rho\right)^{n}\Gamma_{\frac{1-\rho}{1+\rho}}\left(A,B\right)\\
q_{f} & =a^{2}\left(1+\rho\right)^{n}\Gamma_{\frac{1-\rho}{1+\rho}}\left(A\right)\\
q_{g} & =b^{2}\left(1+\rho\right)^{n}\Gamma_{\frac{1-\rho}{1+\rho}}\left(B\right).
\end{align*}
Combining these expressions with \eqref{eq:-4} yields that 
\begin{equation}
q\leq\sqrt{q_{f}q_{g}}.\label{eq:-59}
\end{equation}
Applying the upper bound $\Upsilon_{1}^{\mathtt{UB}}$ respectively
to $q_{f}$ and $q_{g}$, we obtain 
\begin{equation}
q_{f}\le\theta^{+}(a),\qquad q_{g}\le\theta^{+}(b).\label{eq:-60}
\end{equation}
Combining \eqref{eq:-59} and \eqref{eq:-60} yields the desired bound
$\Upsilon_{2}^{\mathtt{UB}}$. 
\end{IEEEproof}
\begin{cor}
\label{cor:symmetric}If $a=b$, then 
\[
\theta^{-}(a)\leq q\leq\theta^{+}(a),
\]
where 
\[
\theta^{-}(t):=\max\left\{ 0,t^{2}-\frac{t}{2}\rho-\left(\frac{t}{2}-t^{2}\right)\rho^{2}\right\} ,
\]
and $\theta^{+}(t)$ is defined in \eqref{eqn:theta_plus}. In particular,
for $a=b=\frac{1}{2}$, 
\begin{equation}
\frac{1-\rho}{4}\le q\leq\frac{1+\rho}{4},\label{eq:-25-1}
\end{equation}
and for $a=b=\frac{1}{4}$, 
\begin{equation}
\frac{1-2\rho-\rho^{2}}{16}\le q\leq\left(\frac{1+\rho}{4}\right)^{2}.\label{eq:-25}
\end{equation}
\end{cor}
The bounds in \eqref{eq:-25-1} are not new; see \cite{witsenhausen1975sequences}.
However, the bounds in \eqref{eq:-25} are new. Note that the bounds
in \eqref{eq:-25-1} are sharp since the upper bound is attained by
the functions $f(\mathbf{x})=g(\mathbf{x})$ and $f(\mathbf{x})=1$
if $x_{1}=1$ and $-1$ otherwise, and the lower bound is attained
by the functions $f(\mathbf{x})=g(-\mathbf{x})$ and $f(\mathbf{x})=1$
if $x_{1}=1$ and $-1$ otherwise.

For $a=b=\frac{1}{4}$, the upper bound in \eqref{eq:-25} is sharp
and attained by the functions $f(\mathbf{x})=g(\mathbf{x})$ and $f(\mathbf{x})=1$
if $x_{1}=x_{2}=1$ and $-1$ otherwise. This answers one of two open
problems posed by Mossel \cite{Mossel2017}. These problems concern
the determination of the maximum $q$ and minimum $q$ when $a=b=\frac{1}{4}$;
we resolve the maximum part of the question. For the minimum $q$,
we do not believe that our lower bound in \eqref{eq:-25} is tight.
In fact, for $a=b=\frac{1}{4}$, we conjecture that $q$ is, in fact,
lower bounded by $\big(\frac{1-\rho}{4}\big)^{2}$. If this conjecture
is true, then this lower bound is sharp since it is attained by the
functions $f(\mathbf{x})=g(-\mathbf{x})$ and $f(\mathbf{x})=1$ if
$x_{1}=x_{2}=1$ and $-1$ otherwise. Furthermore, one may conjecture
that for $a=b=2^{-i},i\in\mathbb{N}$, $\big(\frac{1-\rho}{4}\big)^{i}\le q\leq\big(\frac{1+\rho}{4}\big)^{i}$.
However, this does not hold, since when $a=b$ are small, the $q$
induced by functions $f=g=2\cdot\mathbf{1}_{B}-1$ with $B\subseteq\{-1,1\}^{n}$
denoting a Hamming ball is strictly larger than $\big(\frac{1+\rho}{4}\big)^{i}$,
and moreover, the induced $q$ by such a pair $\left(f,g\right)$
is close to the optimal value \cite[Remark~10.2]{O'Donnell14analysisof}.
A similar conclusion also holds for the case of minimizing $q$.

\subsection{Comparisons to Other Bounds}

Define the maximal correlation between two random variables $X,Y$
as 
\[
\rho_{\mathrm{m}}\left(X;Y\right):=\sup_{f,g}\mathbb{E}\left[f(X)g(Y)\right],
\]
where the supremum is taken over all real-valued Borel-measurable
functions $f$ and $g$ such that $\mathbb{E}\left[f(X)\right]=\mathbb{E}\left[g(Y)\right]=0$
and $\mathbb{E}\left[f^{2}(X)\right]=\mathbb{E}\left[g^{2}(Y)\right]=1$.
When specialized to the binary case, 
\[
\rho_{\mathrm{m}}\left(X;Y\right)=\left|\rho\left(X;Y\right)\right|=\frac{\left|P_{X,Y}(1,1)-P_{X}(1)P_{Y}(1)\right|}{\sqrt{P_{X}(1)P_{X}(-1)P_{Y}(1)P_{Y}(-1)}}.
\]
This quantity was first introduced by Hirschfeld \cite{hirschfeld1935connection}
and Gebelein \cite{gebelein1941statistische}, then studied by R\'enyi
\cite{renyi1959measures}, and it has been exploited to provide a
necessary condition for the NIS problem \cite{witsenhausen1975sequences,kamath2016non}.
Non-interactive simulation of $(U,V)\sim P_{UV}$ using $(\mathbf{X},\mathbf{Y})\sim P_{X,Y}^{n}$
is possible only if $\rho_{\mathrm{m}}\left(U;V\right)\le\rho_{\mathrm{m}}\left(X;Y\right)$.
When specialized to the binary case, it leads to the following bounds
on $q$. 
\begin{prop}[Maximal Correlation Bounds \cite{witsenhausen1975sequences}]
\label{prop:MCB} One has 
\[
ab-\sqrt{a\overline{a}b\overline{b}}\rho\leq q\leq ab+\sqrt{a\overline{a}b\overline{b}}\rho.
\]
\end{prop}
By comparing the maximal correlation upper bound above to $\Upsilon_{1}^{\mathtt{UB}}$,
we get $\Upsilon_{1}^{\mathtt{UB}}\le ab+\sqrt{a\overline{a}b\overline{b}}\rho$.
Equality here holds only when $a=b=\frac{1}{2}$. Hence our upper
bound $\Upsilon_{1}^{\mathtt{UB}}$ is tighter than the maximal correlation
upper bound. On the other hand, comparing the maximal correlation
lower bound above to $\Upsilon_{2}^{\mathtt{LB}}$, we get that for
$a=b$ (i.e., the symmetric case), $\Upsilon_{2}^{\mathtt{LB}}\ge ab-\sqrt{a\overline{a}b\overline{b}}\rho$,
i.e., our bound $\Upsilon_{2}^{\mathtt{LB}}$ is tighter than the
maximal correlation lower bound; for the case $a\le b=\frac{1}{2}$,
$\Upsilon_{2}^{\mathtt{LB}}\leq ab-\sqrt{a\overline{a}b\overline{b}}\rho$,
i.e., our bound $\Upsilon_{2}^{\mathtt{LB}}$ is looser than the maximal
correlation lower bound. Hence, in general, our lower bound $\Upsilon_{2}^{\mathtt{LB}}$
is neither tighter nor looser than the maximal correlation lower bound.

\begin{figure}
\centering %
\begin{tabular}{c}
\includegraphics[width=0.5\textwidth]{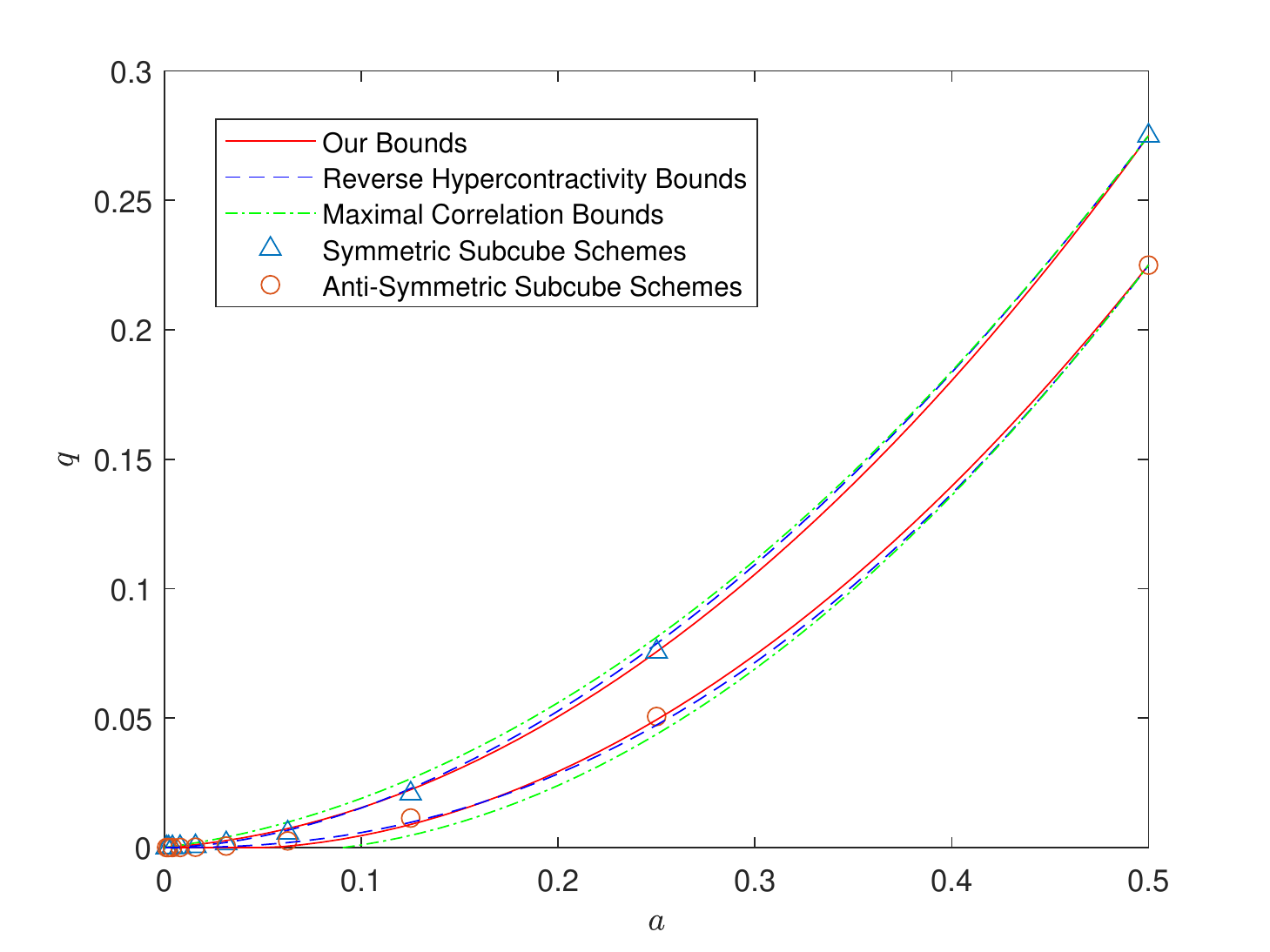} \tabularnewline
{\footnotesize{}{}$\rho=0.1$} \tabularnewline
\includegraphics[width=0.5\textwidth]{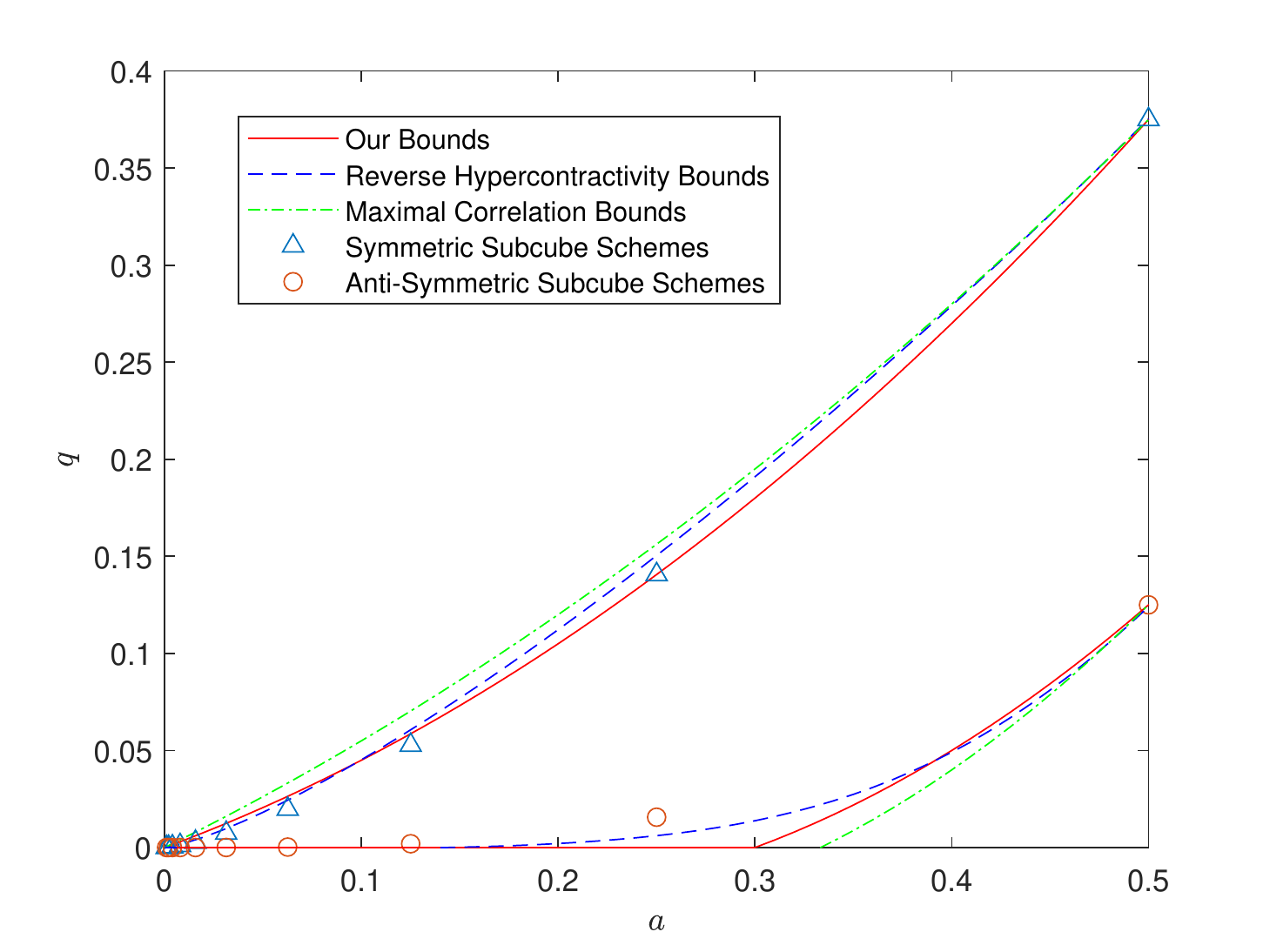} \tabularnewline
{\footnotesize{}{}$\rho=0.5$}\tabularnewline
\multicolumn{1}{c}{\includegraphics[width=0.5\textwidth]{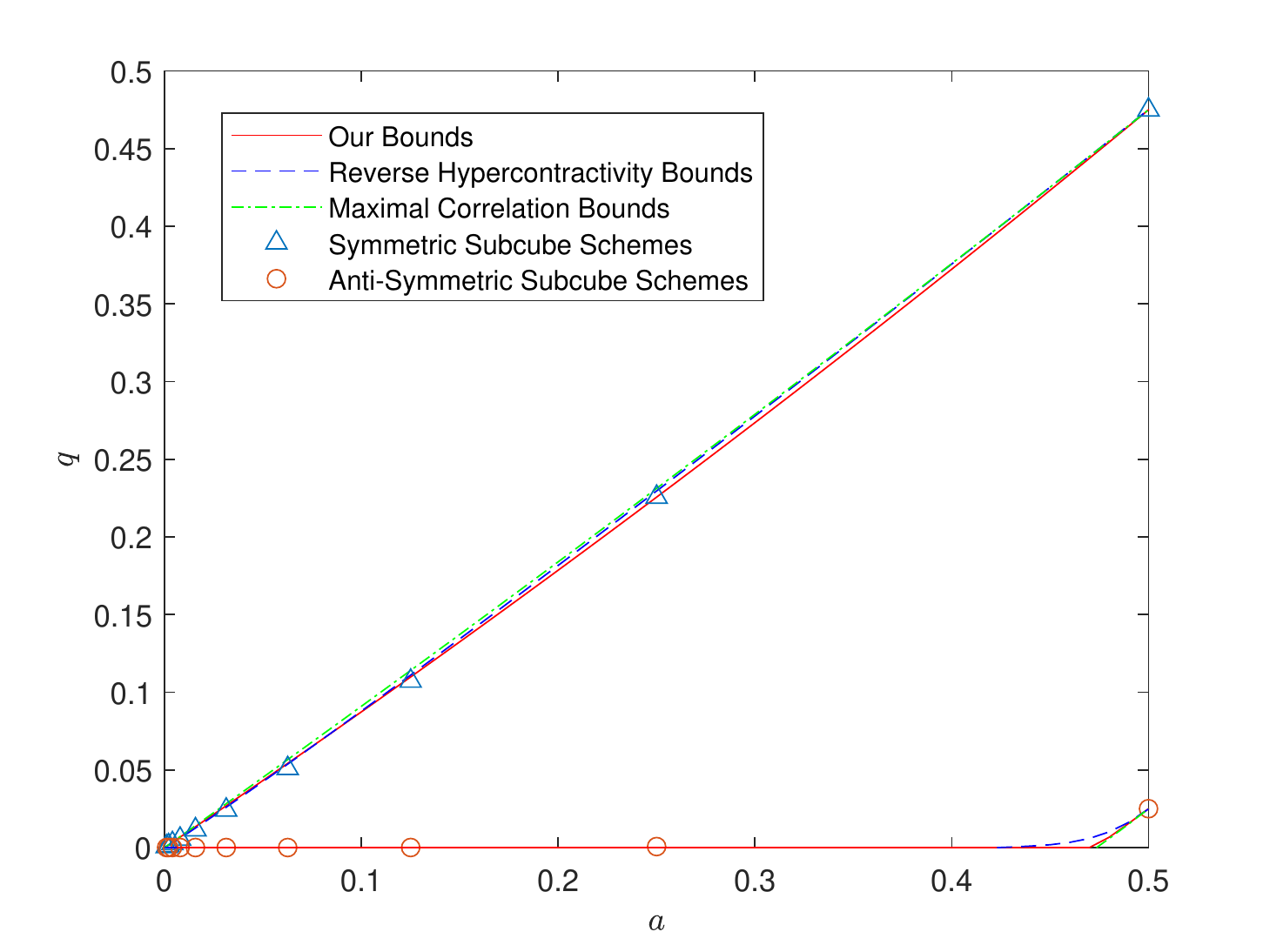} }\tabularnewline
\multicolumn{1}{c}{ {\footnotesize{}{}$\rho=0.9$}}\tabularnewline
\end{tabular}\caption{\label{fig:simulation}Illustration of our bounds in Corollary \ref{cor:symmetric},
the maximal correlation bounds in Proposition \ref{prop:MCB}, and
the hypercontractivity bounds in Proposition \ref{prop:RHB}. %, where
%$a=b$ and the figures from top to bottom respectively correspond
%to the cases of $\rho=0.1$, $0.5$, and $0.9$. 
The upper set of curves correspond to upper bounds, while the lower
set of curves correspond to lower bounds. The symmetric subcube schemes
correspond to the maps $\left\{ \left(f_{i},g_{i}\right)\right\} _{i=1}^{\infty}$
such that $f_{i}(\mathbf{x})=g_{i}(\mathbf{x})$ and $f_{i}(\mathbf{x})=1$
if $x_{1}=x_{2}=\ldots=x_{i}=1$ and $-1$ otherwise. The anti-symmetric
subcube schemes correspond to the maps $\left\{ \left(f_{i},g_{i}\right)\right\} _{i=1}^{\infty}$
such that $f_{i}(\mathbf{x})=g_{i}(-\mathbf{x})$ and $f_{i}(\mathbf{x})=1$
if $x_{1}=x_{2}=\ldots=x_{i}=1$ and $-1$ otherwise. }
\end{figure}

Hypercontractivity (which yields the forward and reverse hypercontractivity
inequalities) is a powerful tool in studying extremal problems, especially
in high-dimensional spaces. Recently, this tool has seen increasing
use for investigating various problems in information theory; see
\cite{O'Donnell14analysisof,anantharam2014hypercontractivity,beigi2015monotone,kamath2016non,liu2018second,raginsky2016strong}
for example. In \cite{mossel2006non,O'Donnell14analysisof,kamath2016non},
Mossel, O'Donnell, Kamath, et al. applied it to derive bounds for
the NIS problem. For example, the following bounds are provided in
\cite{kamath2016non}. More precisely, the following upper bound was
derived by using the forward hypercontractivity inequality, and the
following lower bound by the reverse hypercontractivity inequality.
These are Eqns.~(28) and (29) in~\cite{kamath2016non}. 
\begin{prop}[Hypercontractivity Bounds \cite{kamath2016non}]
\label{prop:RHB} One has 
\begin{align*}
 & \sup_{s,t>0,\left(s-1\right)\left(t-1\right)\left(\kappa-1\right)<0}\varphi_{a,b}\left(s,t,\kappa\right)\\
 & \qquad\leq q\leq\inf_{s,t>0,\left(s-1\right)\left(t-1\right)\left(\kappa-1\right)>0}\varphi_{a,b}\left(s,t,\kappa\right),
\end{align*}
where 
\[
\varphi_{a,b}\left(s,t,\kappa\right):=\frac{\left(s^{\kappa'}a+\overline{a}\right)^{\frac{1}{\kappa'}}\left(t^{\kappa}b+\overline{b}\right)^{\frac{1}{\kappa}}-1}{\left(s-1\right)\left(t-1\right)}-\frac{a}{t-1}-\frac{b}{s-1}
\]
with $\kappa':=1+\frac{\rho^{2}}{\kappa-1}$. 
\end{prop}
Our bounds in Corollary \ref{cor:symmetric}, the maximal correlation
bounds in Proposition \ref{prop:MCB}, and the hypercontractivity
bounds in Proposition \ref{prop:RHB} are plotted in Fig. \ref{fig:simulation}
for the symmetric case in which $a=b$. We set $\rho=0.1$, $0.5$,
and $0.9$. These numerical results show that the hypercontractivity
lower bound is neither tighter nor looser than our lower bound in
general, and the hypercontractivity upper bound is also neither tighter
nor looser than our upper bound in general. More specifically, the
hypercontractivity bounds are tighter than ours for small $a$, while
our bounds are tighter than the hypercontractivity ones for large
$a$ (and they coincide for $a=1/2$). These figures also verify that
both our bounds and the hypercontractivity bounds are uniformly tighter
than the maximal correlation bounds for the symmetric case. The fact
that for the symmetric case, the hypercontractivity bounds are uniformly
tighter than the maximal correlation bounds has been rigorously proven
in \cite[Corollary 1]{kamath2016non}.

\section{Concluding Remarks}

\label{sec:conc}

In the proof of Theorem \ref{thm:DistanceEnumerator}, we applied
Fu-Wei-Yeung's bound \cite{fu2001minimum} on the average distance
(given in \eqref{eq:-6}) to bound the distance enumerator. On the
other hand, as shown in Fig. \ref{fig:simulation}, the hypercontractivity
bounds on the distance enumerator are tighter than our bounds in Theorem
\ref{thm:DistanceEnumerator} in some regimes. Hence, intuitively,
the hypercontractivity bounds can in turn be applied to obtain nontrivial
bounds on the average distance. In fact, the following bound derived
by Chang \cite{chang2002polynomial,O'Donnell14analysisof} is derived
by leveraging such an idea: For $1\le M\le2^{n}$ and $a:=\frac{M}{2^{n}}$,
\begin{equation}
\min_{A:|A|=M}D\left(A\right)\geq\frac{n}{2}-\log\frac{1}{a}.\label{eq:-58}
\end{equation}
Chang's bound was proven by the ``single-function'' version of the
hypercontractivity inequality. Replacing the ``single-function''
version of the hypercontractivity inequality with the ``two-function''
version (i.e., the inequalities in Proposition \ref{prop:RHB}), one
can obtain the following tighter bound (since $\psi\left(a\right)\leq\log\frac{1}{a}$):
For $1\le M\le2^{n}$, we have 
\begin{equation}
\min_{A:|A|=M}D\left(A\right)\geq\frac{n}{2}-\psi\left(a\right),\label{eq:-57}
\end{equation}
where 
\begin{align}
\psi\left(a\right) & :=\inf_{t>0,t\neq1}\frac{\left(ta+\overline{a}\right)\left[at\log t-\left(ta+\overline{a}\right)\log\left(ta+\overline{a}\right)\right]}{a^{2}\left(t-1\right)^{2}}.\label{eq:-45-1}
\end{align}
In Fu-Wei-Yeung's linear programming bound given in \eqref{eq:-6},
$\psi\left(a\right)$ is replaced with $\frac{1}{4a}$. Furthermore,
in \cite{yu2019improved}, the present authors provided an improved
linear programming bound on the average distance. This bound is better
than Fu-Wei-Yeung's bound when $a<\frac{1}{4}$ and is the best over
all existing bounds for $a=\frac{1}{8}$. This improved linear programming
bound was also used to derive an improved bound for the NIS problem
in \cite{yu2019improved}.

In addition, one may wonder whether the hypercontractivity bound for
the NIS problem in Proposition~\ref{prop:RHB} can be improved by
using data processing inequalities (DPIs) arising from the family
of $\Phi$-strong data processing inequality (SDPI) constants~\cite[Theorem~19]{beigi2018phi}
or $\Phi$-ribbons~\cite[Theorem~12]{beigi2018phi} which were proposed
by Beigi and Gohari~\cite{beigi2018phi}. We term bounds resulting
from these two DPIs as the $\Phi$-SDPI bound and the $\Phi$-ribbon
bound respectively. Then, Kamath-Anantharam's hypercontractivity upper
bound in Proposition~\ref{prop:RHB} is a special case of the $\Phi$-ribbon
bound with $\Phi\left(t\right)=t\log t$~\cite{beigi2018phi}. We
conjecture that in our setting (i.e., DSBS $(X,Y)$ and binary $U$
and $V$), under the common tensorization condition that $\Phi\in\mathscr{F}$
with the family of functions $\mathscr{F}$ defined in~\cite[Definition~6]{beigi2018phi},
the hypercontractivity upper bound in Proposition~\ref{prop:RHB}
is the tightest over all $\Phi$-ribbon bounds. In fact, the first
author of this paper and V.~Anantharam have shown a weaker version
of this conjecture. That is, the bound derived from the DPI $s^{*}(U,V)\leq s^{*}(X,Y)$
with $s^{*}(W,Z)$ denoting the hypercontractivity constant of $(W,Z)$,
is the tightest $\Phi$-SDPI bound over all $\Phi\in\mathscr{F}$~\cite{yu2020hypercontractivity}.
This will be the content of a future paper~\cite{yu2020hypercontractivity}.

\subsection*{Acknowledgements}

The authors thank the associate
editor Prof.\ Amin Gohari and the two reviewers for insightful comments
that have helped to improve the paper.  \bibliographystyle{unsrt}
\bibliography{ref}
\begin{IEEEbiographynophoto}{Lei Yu} received the B.E. and Ph.D. degrees, both in electronic engineering, from University of Science and Technology of China (USTC) in 2010 and 2015, respectively. From 2015 to 2017, he was a postdoctoral researcher at the Department of Electronic Engineering and Information Science (EEIS), USTC. From 2017 to 2019, he was a research fellow at the Department of Electrical and Computer Engineering, National University of Singapore. Currently, he is a postdoc at the Department of Electrical Engineering and Computer Sciences, University of California, Berkeley. His research interests lie in the intersection of information theory, probability theory, and combinatorics.  \end{IEEEbiographynophoto}

\begin{IEEEbiographynophoto}
{Vincent Y.\ F.\ Tan} (S'07-M'11-SM'15)   was born in Singapore in 1981. He is currently a Dean's Chair Associate Professor in the Department of Electrical and Computer Engineering  and the Department of Mathematics at the National University of Singapore (NUS). He received the B.A.\ and M.Eng.\ degrees in Electrical and Information Sciences from Cambridge University in 2005 and the Ph.D.\ degree in Electrical Engineering and Computer Science (EECS) from the Massachusetts Institute of Technology (MIT)  in 2011.  His research interests include information theory, machine learning, and statistical signal processing.

Dr.\ Tan received the MIT EECS Jin-Au Kong outstanding doctoral thesis prize in 2011, the NUS Young Investigator Award in 2014,  the Singapore National Research Foundation (NRF) Fellowship (Class of 2018) and the NUS Young Researcher Award in 2019. He was also an IEEE Information Theory Society Distinguished Lecturer for 2018/9. He is currently serving as an Associate Editor of the {\em IEEE Transactions on Signal Processing} and an Associate Editor of Machine Learning for the {\em IEEE Transactions on Information Theory}. He is a member of the IEEE Information Theory Society Board of Governors. 
\end{IEEEbiographynophoto} 
%{Vincent Y.\ F.\ Tan} (S'07-M'11-SM'15) was born in Singapore in 1981. He is currently a Dean's Chair Associate Professor in the Department of Electrical and Computer Engineering and the Department of Mathematics at the National University of Singapore (NUS). He received the B.A.\ and M.Eng.\ degrees in Electrical and Information Sciences from Cambridge University in 2005 and the Ph.D.\ degree in Electrical Engineering and Computer Science (EECS) from the Massachusetts Institute of Technology (MIT) in 2011. His research interests include information theory, machine learning, and statistical signal processing.
%Dr.\ Tan received the MIT EECS Jin-Au Kong outstanding doctoral thesis prize in 2011, the NUS Young Investigator Award in 2014, the NUS Engineering Young Researcher Award in 2018, and the Singapore National Research Foundation (NRF) Fellowship (Class of 2018). He is also an IEEE Information Theory Society Distinguished Lecturer for 2018/9. He has authored a research monograph on {\em ``Asymptotic Estimates in Information Theory with Non-Vanishing Error Probabilities''} in the Foundations and Trends in Communications and Information Theory Series (NOW Publishers). He is currently serving as an Associate Editor of the IEEE Transactions on Signal Processing.  

\end{document}